\documentclass{epl}

\title{Metamagnetism of itinerant electrons in multi-layer ruthenates}
\shorttitle{Metamagnetism in multi-layer ruthenates}
\author{Benedikt Binz\inst{1,2} \and Manfred Sigrist\inst{1}}
\institute{
  \inst{1}D\'epartement de physique, Universit\'e de Fribourg, P\'erolles,
  CH-1700 Fribourg, Switzerland \\
  \inst{2}Theoretische Physik, ETH-H\"onggerberg, CH-8093 Z\"urich,
  Switzerland}

\begin{document}

\def\up{\uparrow}
\def\down{\downarrow}
\newcommand{\eref}[1]{(\ref{#1})}

\maketitle

\begin{abstract}
  The problem of quantum criticality in the context of itinerant ferro- or
  metamagnetism has received considerable attention [S. A. Grigera {\it et.
    al.}, Science {\bf 294}, 329 (2001); C. Pfleiderer {\it et. al.}, Nature,
  {\bf 414}, 427 (2001)]. It has been proposed that a new kind of quantum
  criticality is realised in materials such as \chem{MnSi} or
  \chem{Sr_3Ru_2O_7}. We show 
  based on a mean-field theory that the low-temperature behaviour of the
  \mth{n}-layer ruthenates \chem{Sr_{n+1}Ru_nO_{3n+1}} can be understood as a
  result of a Van Hove singularity (\acro{VHS}). We consider a single band whose Fermi energy, \mth{E_F}, is close to the \acro{VHS} and deduce a complex phase diagram for
  the magnetism as a function of temperature, magnetic field and \mth{E_F}. The
  location of \mth{E_F} with respect to the \acro{VHS} depends on the number of layers or
  can be tuned by pressure. We find that the ferromagnetic quantum phase
  transition in this case is not of second but of first order, with a metamagnetic
  quantum critical endpoint at high magnetic field. Despite its simplicity
  this model describes well the properties of the uniform magnetism in the
  single, double and triple layer ruthenates. We would like to emphasise that
  the origin of this behaviour lies in the band structure.
\end{abstract}

\section{Introduction}

The issue of metamagnetism in itinerant electron systems was studied both 
theoretically 
\cite{wohlfarth62,shimizu82} and experimentally \cite{goto89} long ago.
Recently, the investigation of metamagnetism in two compounds, \chem{MnSi}
\cite{pfleiderer01} and \chem{Sr_3Ru_2O_7} \cite{grigera01,grigera02},
has revived interest in this phenomenon,  considering it from a
new point of view. 
It has been suggested that these systems might display 
a new type of quantum criticality, connected with a so-called quantum critical
end point (\acro{QCEP}), in the vicinity of which the Landau Fermi-liquid
theory of metals breaks down. In fact the field \mth{H_m(T)} initiating
the metamagnetic transition, the abrupt increase of the magnetisation, 
defines a line of first-order transitions in the
field (H)-temperature (T) plane without any symmetry-breaking. The
first-order line ends in a critical end point \mth{(H_c,T_c)}. A \acro{QCEP}
occurs if \mth{T_c} is suppressed to zero as a function of an additional parameter,
such as the pressure or the chemical composition of the material. 

While theoretical results concerning the properties of the \acro{QCEP}
have been obtained recently on the basis of phenomenological, low-energy field
theories \cite{millis02}, to our knowledge no discussion of the
microscopic origin of such a \acro{QCEP} has yet been provided.
In the following, we present a model which gives rise to a \acro{QCEP} of the
above kind, which we will discuss within a simple Hartree-Fock theory. 
This model is based on a system with  a single,
two-dimensional electron band where the Fermi 
level is close to a Van Hove singularity (\acro{VHS}). Our
results for the magnetisation as a function of field, temperature and the band
filling  are in qualitative agreement with observed properties of the two- and
three-layer ruthenate compounds \cite{perry01,cao03}, and suggest that
the general magnetic phase diagram of \mth{n}-layer ruthenates might be
understood in terms of band-structure properties. 

\section{Mean-field theory of itinerant  uniform magnetism}

We consider a single-band model of electrons interacting via an on-site
Coulomb repulsion \mth{U}. In the Hartree-Fock (or mean-field) approximation, the
free energy per unit cell \mth{f} is given as a function of the particle density
\mth{n}, the uniform magnetisation \mth{m} (in units of \mth{g\mu_B} per unit cell) and the
temperature by 
\begin{equation} 
f_\ab{HF}(n,m,T)=f_0(n,m,T)+U n_\up n_\down, 
\end{equation} 
where
\mth{n_{\up,\down}=n/2\pm m} are the densities of up- and down-spin electrons,
respectively and \mth{f_0} is the free energy in the absence of interactions, {\it i.e.}
\begin{equation} 
f_0(n,m,T)=\sum_{\sigma=\up,\down} \left[\Omega_0(\mu_\sigma,T)+\mu_\sigma
  n_\sigma\right], 
\end{equation} 
where \mth{\Omega_0(\mu_\sigma,T)} is the grand-canonical
potential and \mth{\mu_\sigma(n_\sigma,T)} are the chemical potentials for the two
species of electrons in the absence of interactions. The only
information about the band structure which enters the mean-field
theory is the
density of states (\acro{DOS}) as a function of energy,
\mth{\rho(\omega)=1/V\sum_{\vect{k}}\delta(\omega-\varepsilon_{\vect{k}})},
where \mth{\varepsilon_{\vect{k}}} is the 
electron dispersion and the sum is over the first Brillouin
zone. Given \mth{\rho(\omega)}, 
one obtains \mth{\Omega_0(\mu_\sigma,T)=-k_B T\int\!\upd\varepsilon\,\rho(\varepsilon)
\ln{\left(1+e^{-\beta(\varepsilon-\mu_\sigma)}\right)}}, where \mth{k_B} is the
Boltzmann constant and \mth{\beta=(k_B T)^{-1}}. The chemical potentials
\mth{\mu_\sigma(n_\sigma,T)} are given implicitly by
\begin{equation}
n_\sigma=-\partial_{\mu_\sigma}\Omega_0=\int\!\upd\varepsilon\,
\rho(\varepsilon)f_T(\varepsilon-\mu_\sigma),  
\end{equation} 
where \mth{f_T(\varepsilon)=1/(\exp(\beta\varepsilon)+1)} is the Fermi function.

In the following, we omit the suffix ``HF'' in \mth{f_\ab{HF}}. Once
the free energy is determined, we obtain the thermodynamic equation of state
\begin{equation} 
h=\partial_mf=\mu_\up-\mu_\down-2Um,\label{h} 
\end{equation} 
where \mth{h} is  the external magnetic field multiplied by \mth{g\mu_B}. 
Equation \eref{h} may be solved for \mth{m} to
obtain the magnetisation as a function of \mth{n,T} and \mth{h}. The thermodynamically
stable solution minimises the Gibbs free energy, \mth{g=f-hm}.

A first-order transition is characterised by a
discontinuous jump of the magnetisation, whereas a smooth metamagnetic
transition occurs if the differential susceptibility \mth{\chi=\partial m/\partial
h} has a (more or less pronounced) maximum. The susceptibility is obtained
from 
\begin{equation}
\frac1{\chi}=\partial_m^2f=\sum_\sigma\frac1{A_0(\mu_\sigma,T)}-2U,\label{chi}
\end{equation} 
where
\mth{A_n(\mu,T)=\int\!\upd\varepsilon\,\rho^{(n+1)}(\varepsilon)
f_T(\varepsilon-\mu)}. 
Note that \mth{\lim_{T\to0}A_n(\mu,T)=\rho^{(n)}(\mu)}, where \mth{\rho^{(n)}(\mu)} denotes the \mth{n}-th derivative of
the \acro{DOS} at the Fermi level. Because \mth{\partial_m\chi|_{m=0}=0} and
\mth{\lim_{m\to\infty}\chi=0}, a sufficient (but not necessary) condition for
either smooth or first-order metamagnetism is given by
\mth{\partial_m^2\chi|_{m=0}>0}, which is equivalent to \mth{\partial_m^4f|_{m=0}<0}
or \mth{A_0 A_2>3(A_1)^2}, where \mth{A_n} is evaluated at
\mth{\mu=\mu_\up=\mu_\down}, {\it i.e.} in
the absence of magnetisation. At zero temperature, this is equivalent to the condition
\begin{equation}
\rho\rho''>3(\rho')^2,\label{mmcond} 
\end{equation} 
which has been  discussed by Wohlfarth and Rhodes\cite{wohlfarth62}. 
The condition is clearly satisfied
if the \acro{DOS} is large and has a strong positive curvature at the Fermi level.

Such a positive curvature also gives rise to two unusual finite-temperature
properties. First, \mth{\chi(T)|_{m=0}} is not monotonous, but has a maximum as a
function of \mth{T}. Second, the entropy \mth{s=-\partial_Tf} at low
temperatures does not decrease monotonously with the magnetisation, but has a maximum at a finite \mth{m}. Both
phenomena are related to the fact that, if 
\mth{\rho(\varepsilon)} is positively curved, the average \acro{DOS} in a
finite region around \mth{\varepsilon_F} is higher than the \acro{DOS}
at the Fermi level. A sufficient condition for the two properties mentioned
above is given by \mth{\partial_T^2\partial_m^2f|_{m=T=0}<0}
or \mth{\rho\rho''>(\rho')^2}, which is a weaker condition than eq.~\eref{mmcond}.

Equation \eref{mmcond} is satisfied if, for example, the Fermi level
occurs at a local minimum of the \acro{DOS}. This case probably applies to
\chem{MnSi} \cite{yamada99}. Another example, which we wish to address 
here, occurs if the Fermi level is close to a \acro{VHS}. This is the case
for the \chem{Sr}-ruthenates which as layered perovskite systems
represent quasi-two-dimensional electron systems. In two dimensions 
the \acro{VHS} is logarithmic, {\it
  i.e.} the \acro{DOS} has the asymptotic behaviour \mth{\rho(\varepsilon)=\frac1{W_1}\ln{|\frac{W_2}\varepsilon|}}, where
\mth{W_{1,2}} are two parameters with units of energy, which are both of
the order of the bandwidth\footnote{As an example, the \mth{\gamma}-band of \chem{Sr_2RuO_4} 
is well approximated by  a two-dimensional tight-binding model with nearest-neighbour hopping 
\mth{t=0.17\,eV} and diagonal next-nearest-neighbour hopping \mth{t'=0.08\,eV}. For this model 
we find \mth{W_1=2\pi^2\sqrt{t^2-4t'^2}\approx1.14\,eV} and \mth{W_2\approx 0.92\,eV}.}. For
simplicity, we choose \mth{W_1=W_2=W} in the following, and we extend the logarithmic 
form of the \acro{DOS} from \mth{\varepsilon=-W} to \mth{+W}.

One way to study phase transitions is to develop \mth{f} in powers of
\mth{m}. This has been performed previously for related systems \cite{wohlfarth62,shimizu82,yamada99}, but the application of such a
procedure can be complicated. At least at zero \mth{T}, this approach is not successful in the present case, 
because as soon as eq.~\eref{mmcond} is satisfied,
\mth{\partial_m^{n}f|_{m=T=0}\leq0} for all 
\mth{n\geq4}. In fact, \mth{f(m)|_{T=0}} is not analytic at the specific value
of \mth{m} where one of 
the Fermi levels crosses the \acro{VHS}.

\section{Results}

\begin{figure}
  \twofigures[scale=0.5]{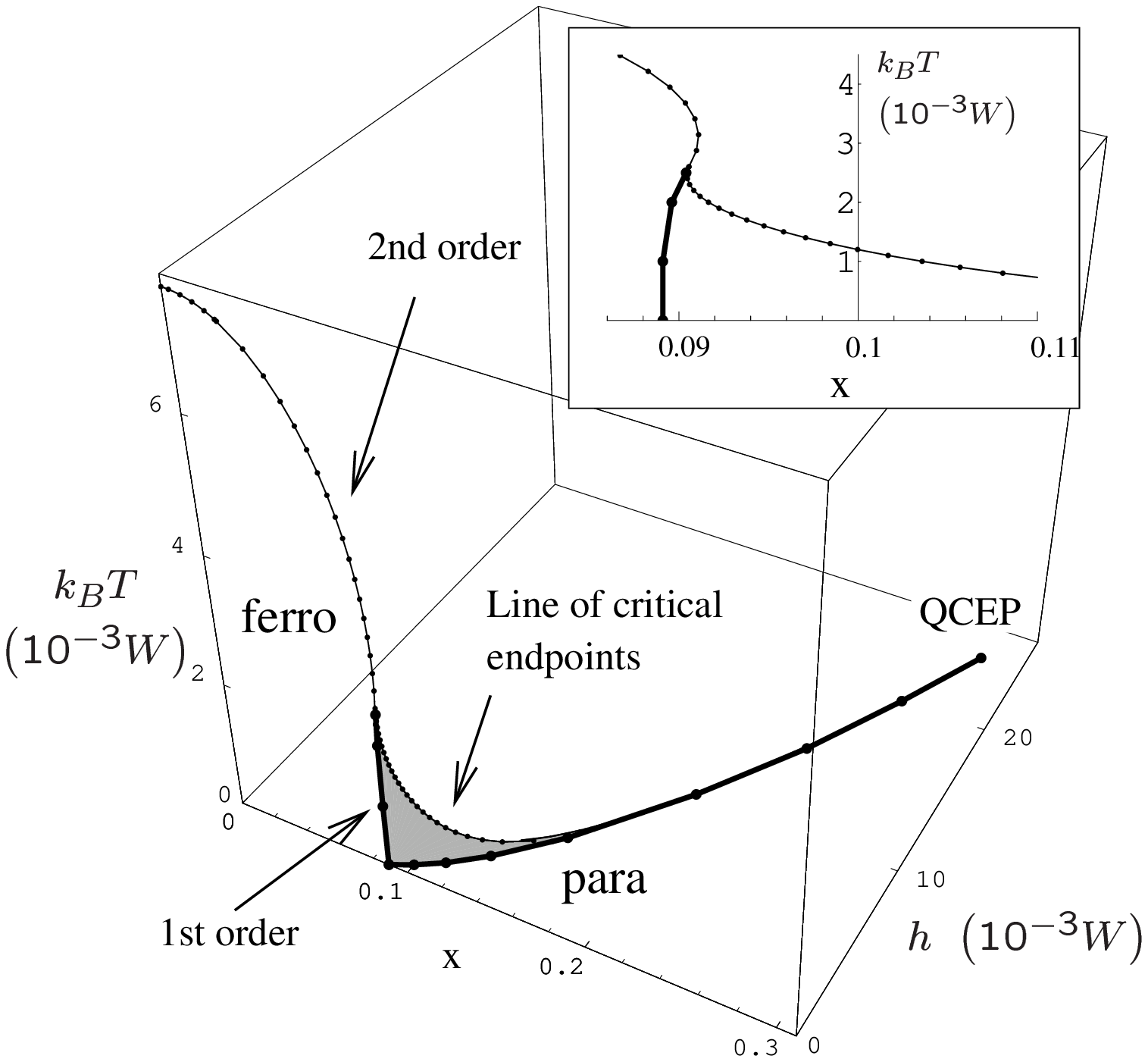}{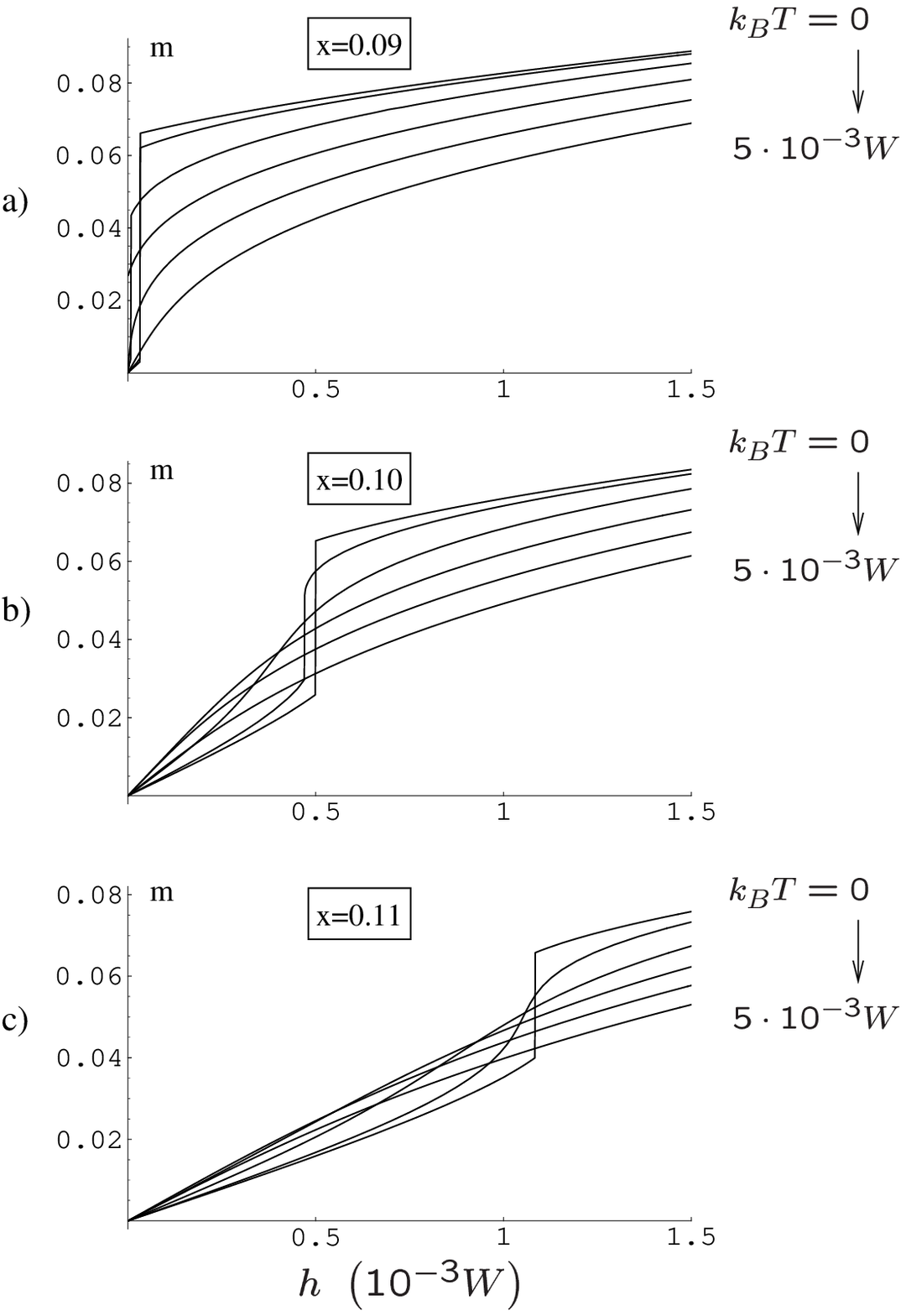}
\caption{The phase diagram for \mth{U=0.2W} as a function of temperature,
  magnetic field and
  \mth{x}, the separation of the electron density from the Van Hove filling. The
  transition from ferro- to paramagnetic behaviour changes from second to first
  order. The grey surface indicates a first-order metamagnetic transition
  ({\it i.e.} a jump in the magnetisation) and terminates at a line of critical
  endpoints. This line is suppressed rapidly to low temperatures and finally
  reaches zero temperature at a \acro{QCEP}. The inset shows a
detailed section of the phase diagram projected on the \mth{(x,T)}-plane.}
  \label{phasediag}
\caption{Magnetisation as a function of external field for three different fillings and for
 temperatures \mth{k_BT=0}, \mth{1}, \mth{2}, \mth{3}, \mth{4}, \mth{5\cdot 10^{-3}W}.} 
\label{mofh}
\end{figure}

Despite its simplicity, our model yields a complex mean-field phase
diagram. In
fig.~\ref{phasediag}, we
show the phase diagram as a function of temperature, magnetic field and the
parameter \mth{x}, which is the difference between the actual electron
density and the density at Van Hove filling, {\it i.e.}  \mth{x=n-2}.

If the system is very close to the \acro{VHS}, we find the usual
Stoner instability, {\it i.e.} a second-order transition as a function of
temperature. The line of second-order transitions within the \mth{(T,x)}-plane of
the phase diagram is characterised by a diverging susceptibility, or
equivalently \mth{UA_0(\mu(x,T),T)=1}.  
If the system is pushed further away from the \acro{VHS},
the simple picture of a ferromagnetic low-temperature and a 
paramagnetic high-temperature phase is no longer valid, and we find a weak
reentrant behaviour, {\it i.e.} the system is ferromagnetic for intermediate
temperatures, 
\mth{T_{1}<T<T_{2}}. The reason for this behaviour lies in the entropy
anomaly discussed above. The reentrant 
behaviour occurs only in a very narrow range of \mth{x} 
(\mth{0.089<x<0.091} for \mth{U=0.2W}).

At low temperatures, the (reentrant) transition between the ferro- and the
paramagnetic phase is of first order, {\it i.e.} it occurs due to the crossing of two minima of \mth{g(m)=f(m)-hm}. For example, the quantum phase
transition as a function of \mth{x} at 
\mth{T=h=0} occurs at \mth{x\approx0.089}, whereas the Stoner criterion
\mth{U\rho(\varepsilon_F)=1} would give a slightly smaller value, \mth{x\approx0.081}. A metastable ferromagnetic state, which is present
in the paramagnetic region close to the first-order transition, 
gives rise to a first-order 
transition in an external magnetic field. This transition is 
is characterised by a jump of the magnetisation without any
spontaneous symmetry-breaking. The critical field 
of the transition {\it decreases} slightly with temperature
at fixed values of \mth{x}.

The line of first-order transitions (at fixed \mth{x}) terminates at a
critical endpoint 
\mth{(h_c(x),T_c(x))}, which forms a line of critical endpoints in the
parameter space including \mth{x}. This line of critical endpoints is
determined by the  
equations:
\mth{\partial_m^2 f=\partial_m^3 f=0} and eq.~\eref{h}, which must be
solved for  \mth{h}, \mth{T} and \mth{m} at a given value of \mth{x}. The second derivative is
given in eq.~\eref{chi} and an explicit expression for the third
derivative is \mth{\partial_m^3
f=A_1(\mu_\down)/A^3_0(\mu_\down)-A_1(\mu_\up)/A^3_0(\mu_\up)}.  


With increasing separation from the Van Hove filling, the
critical point moves rapidly to lower temperatures and higher fields. Finally,
the line of critical endpoints reaches zero temperature, thus giving rise to a
\acro{QCEP} at the point \mth{(x_{qc},h_{qc})}. The low-temperature limit is more
involved. In fact, the equation \mth{\partial_m^3f=0} must be 
replaced at zero temperature by the requirement that one of the two
spin bands, for example the spin-down band, is at Van Hove filling,
{\it i.e.} \mth{\mu_\down|_{\ab{QCEP}}=0},
which implies that \mth{x_{qc}=2m_{qc}}.
As a consequence \mth{\partial_m^3 f} is not zero at
the \acro{QCEP} and actually has a pole. We find
\begin{eqnarray}
x_{qc}&=&(1+\frac W{2U})e^{-\frac W{2U}},\\
2h_{qc}&=&(W-2U)e^{-\frac W{2U}},
\end{eqnarray}
where \mth{W} is the energy parameter in the model \acro{DOS}. For \mth{U=0.2 \,W}, 
this yields \mth{x_{qc}\approx0.29} and
\mth{h_{qc}\approx25\cdot 10^{-3}W}. The basic structure of the phase diagram of
fig.~\ref{phasediag} is invariant inside the range \mth{0<U<W/2},
although the details depend strongly on \mth{U}. For \mth{U>W/2}, both first
order metamagnetism and quantum 
criticality disappear. 

Within our
model and choice of parameters, the \acro{QCEP} lies outside the 
experimentally accessible region. Accessing this point would require a magnetic
field in the order 
of 200 Tesla, and even if such enormous
fields could be created, the metamagnetic 
behaviour ({\it i.e.} the anomaly in \mth{m(h)}) is negligible. We find
that \mth{T_c} is suppressed exponentially as the 
\acro{QCEP} is approached, {\it i.e.} \mth{T_c(x)\sim\exp(\alpha/(x-x_{qc}))} with
\mth{\alpha=\left(\frac W{2U}\right)^3\exp(-\frac W{2U})\approx1.28}. This
leads to critical endpoints at 
extremely low temperatures for realistic values of the magnetic field,
which are practically indistinguishable from the true \acro{QCEP}. The Fermi
level, \mth{\mu_\down}, at the critical end point vanishes even faster than
\mth{T_c} for \mth{x\to x_{qc}}.  

In fig.~\ref{mofh}, we show the equilibrium magnetisation as a function of the
magnetic field for three different fillings and for various temperatures. At
\mth{x=0.09}, the system is in the reentrant ferromagnetic regime. At very low
temperatures, a first-order magnetic transition can be found at a
small but finite 
field. A metastable magnetic phase also exists in the absence of the
external field. True equilibrium ferromagnetism is established only at
intermediate temperatures, and there is a second-order phase transition
towards a high-temperature paramagnetic state. 

 For \mth{x=0.1},  the system is in the metamagnetic
regime. At low temperatures there is a first-order metamagnetic
transition. Both the critical field of this transition and the size of the
magnetisation jump are decreasing functions of the
temperature.  The discontinuity is transformed
into a smooth metamagnetic crossover above the critical temperature,
\mth{k_BT_c\approx1.2\cdot 10^{-3}W}, and at still higher temperatures, the normal paramagnetic
behaviour is recovered. Further away from the Van Hove filling, at \mth{x=0.11}
(fig.~\ref{mofh} c)), the 
behaviour is similar, but the metamagnetic transition occurs at higher fields
and \mth{T_c} is reduced.

\begin{figure}
  \onefigure[scale=0.5]{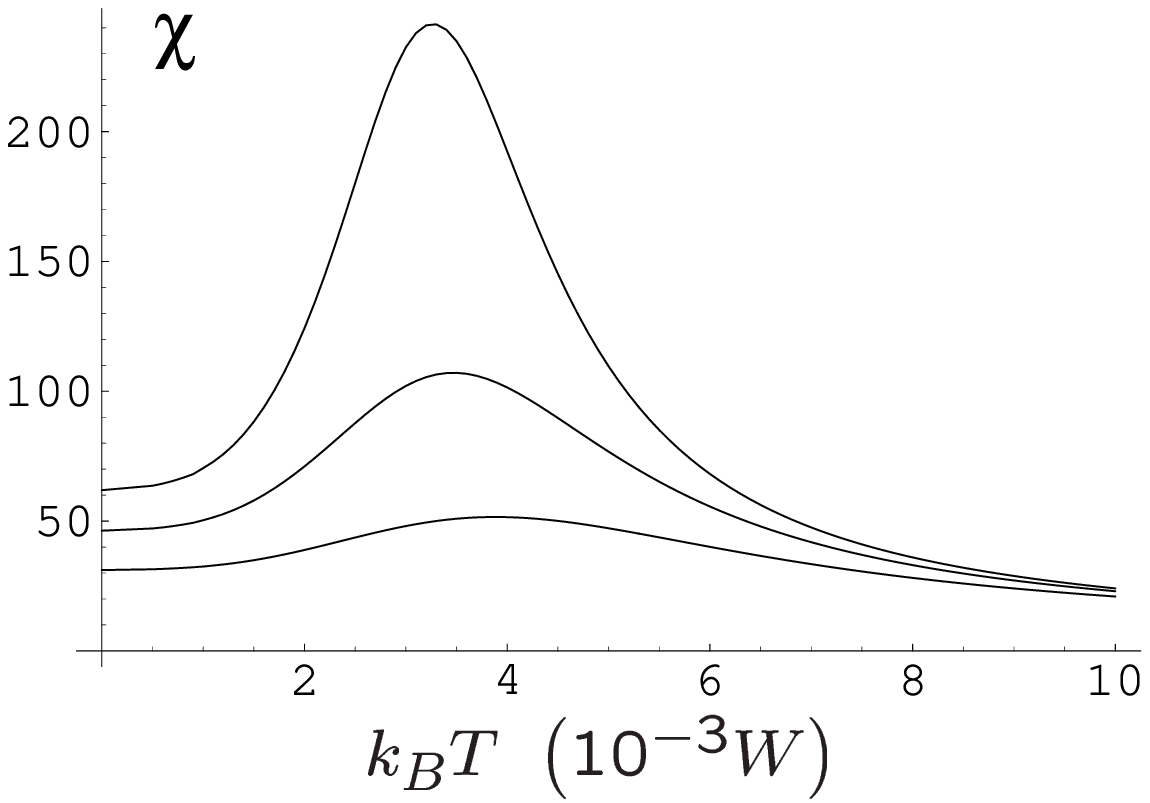}
\caption{Zero-field spin susceptibility as a function of temperature for
  three different fillings, \mth{x=0.095}, \mth{0.1} and \mth{0.11} (from top to bottom).} 
\label{chiofT}
\end{figure}

From figs.~\ref{mofh} b) and c), we observe that the linear response to
low fields is not a monotonous function of 
temperature,  but has a maximum at a finite temperature \mth{T^*} which is higher
than \mth{T_c}. For fillings close to the reentrant ferromagnetic regime, 
\mth{k_BT^*} approaches \mth{3.2\cdot 10^{-3}W}, the temperature where ferromagnetism first appears, as 
\mth{x} is lowered. At this point, which corresponds to a second-order ferromagnetic transition, the susceptibility is in fact a diverging quantity. For larger values
of \mth{x}, the peak in \mth{\chi(T)} is much less pronounced and \mth{T^*} increases (fig.~\ref{chiofT}). A very broad maximum remains even for
fillings much larger than \mth{x_{qc}}.

\section{Comparison with experiment}

In view of the metamagnetic behaviour of the multi-layer
strontium-ruthenates (the layered perovskite compounds \chem{Sr_{1+n}Ru_nO_{3n+1}}), it is natural to 
concentrate on the electronic 
band which is related to the \mth{ \gamma }-band of the single-layer
compound. This band, which originates from the 4\mth{d_{xy}} orbitals of \chem{Ru}, lies
close to a \acro{VHS}. The double-layer compound \chem{Sr_{3}Ru_2O_{7}}
has a modified band structure which preserves two bands with the basic \mth{ \gamma }-band
structure. Band structure calculations suggest that one band is moved closer to the \acro{VHS} \cite{hase96,singh95}. It is natural to assume
that this trend of approaching the \acro{VHS} in one band progresses with
increasing number \mth{n} of layers in each unit cell, so that \mth{ n} is a
tuning parameter for the variable \mth{x} in one of the bands. 
Other aspects, such as a possible change in the ratio
between the bandwidth and the Coulomb repulsion or more complex
multi-band physics, are neglected. In fact a phase diagram very
similar to  fig.~\ref{phasediag}, where the filling parameter \mth{x} is
replaced by the pressure, was proposed on purely phenomenological
grounds in relation to \chem{Sr_3Ru_2O_7} \cite{grigera02}.  
For a given compound, 
\mth{x} may also be tuned by external pressure or by chemical pressure,
such as the gradual replacement of Sr by the smaller isoelectronic ion
Ca. 

The magnetisation curves in fig.~\ref{mofh} b) are in good qualitative
agreement with the 
observed behaviour in \chem{Sr_4Ru_3O_{10}} \cite{cao03} for magnetic
fields parallel to the 
\mth{ab}-plane\footnote{\chem{Sr_4Ru_3O_{10}} is actually ferromagnetic
  below \mth{105K}, with an easy axis in the \mth{c}-direction. Metamagnetic
  behaviour was observed by applying a field parallel to the
  \mth{ab}-plane. 
The very different magnetic behaviour parallel and
perpendicular to the \mth{c}-axis is an effect of spin-orbit coupling,
which is not included in our model.}. Furthermore, fig.~\ref{mofh} c) is
 consistent with the experimental 
findings in \chem{Sr_3Ru_2O_7}\cite{perry01}, where smooth
metamagnetism has been 
observed at fields which are 2-3 times larger than the critical field in \chem{Sr_4Ru_3O_{10}} \cite{cao03}.
We would like to emphasise that
the basic, qualitative properties of the double- and triple-layer
compounds are in good agreement with our model calculation. Our
choice of parameters even yields values of 
 temperature and magnetic field which are of the same order of magnitude as
 those found in experiment. 
From our calculations, we
expect that a first-order metamagnetic transition  should be found at sufficiently low
temperatures in  \chem{Sr_3Ru_2O_7}, but given the rapid decrease of \mth{T_c} as a
function of \mth{x}, the required temperature is likely to be much smaller than that attained in the experiment.

We stress that in our model the \acro{QCEP} occurs when one
band is at the Van Hove filling. This 
circumstance may influence unconventional
transport properties in the vicinity of the \acro{QCEP} \cite{grigera01}, for non-Fermi-liquid behaviour has been suggested as one consequence of an itinerant electron band being at the
Van Hove filling \cite{newns94,dzyaloshinskii96,irkhin01}. 

Instead of increasing the number of layers, an alternative approach to
the \acro{VHS} may be through electron doping of the single-layer
compound.   This was attempted recently by partial
replacement of \chem{Sr} by \chem{La} \cite{kikugawa02}. The experimental
result indeed shows an enhanced
magnetic response as a function of doping, but neither ferromagnetism nor the typical temperature dependence of the susceptibility, with
a maximum at finite temperature, was observed.
The question of the extend to which the doping may lead to an
increase in the itinerant electron density remains to be answered.
 If the additional electrons
remain essentially localised around the dopants, this may yield a
similarly enhanced, almost Curie-like, susceptibility \cite{kikugawa02}.

\section{Conclusion}

In conclusion, we have presented a mean-field theory of uniform magnetism for
a single band of itinerant electrons which is close to a \acro{VHS}. The
resulting phase diagram as shown in fig.~\ref{phasediag}, 
features first- and second-order ferromagnetic transitions, as well as a 
line of (metamagnetic) critical endpoints, which is pushed to zero temperature
at a \acro{QCEP}. Such a phase diagram has been proposed on purely phenomenological
grounds in the context of \chem{MnSi}\cite{pfleiderer01} and multi-layer
ruthenates \cite{grigera02}, but no microscopic description has yet been provided. Despite its simplicity our model calculation gives a good
qualitative description of the magnetic properties of two- and
three-layer ruthenates 
\cite{perry01,cao03}. The general magnetic behaviour of these materials may
thus be explained in terms of band-structure properties.  A similar scenario
  applies to \chem{MnSi}, where the Fermi energy is likely to be associated
  with a dip in
  the electron DOS, and to \chem{FeSi_{1-x}Ge_x}, which exhibits a transition 
  from a small-gap semiconductor to a metallic ferromagnet with increasing
  \mth{x}\cite{anisimov02}. 

Our analysis reveals the remarkable aspect that
the \acro{QCEP} is characterised by the situation in which the chemical
potential  of one of the bands (minority- or majority-spin) is exactly at the 
Van Hove
filling. This should have non-trivial consequences for the metallic properties
close to the \acro{QCEP}. In particular, transport properties may show
non-Fermi-liquid behaviour \cite{newns94,irkhin01}, similar
to the experimental observation \cite{grigera01}.

Finally, we comment that, although the model we have presented captures many of the qualitiative features observed in multi-layer ruthenates, it remains a relatively crude simplification. Here we have 
neglected several aspects, including multi-band effects, spin-orbit coupling and disorder effects, which undoubtedly have a certain role in real
materials. One of the most important questions concerns the effects of spin fluctuations,
which are neglected within the Hartree-Fock approximation. These aspects are
the subject of current studies.

\acknowledgments
We thank H.~B.~Braun, A.~Mackenzie, Y.~Maeno, T.~M.~Rice and  C.~Stafford for helpful and stimulating discussions and B.~Normand for carefully proof-reading our manuscript. This work was supported by the Swiss National Science Foundation.

\end{document}